\documentclass[letter]{aa} 

\usepackage{graphicx}
\usepackage{txfonts}
\begin{document}

   \title{Larger $\lambda_R$ in the disc of isolated active spiral galaxies than in their non-active twins}

   \author{I. del Moral-Castro\inst{\ref{inst1}\and\ref{inst2}}\thanks{E-mail: ignaciodelmoralcastro@gmail.com}
          \and
          B. Garc\'ia-Lorenzo\inst{\ref{inst1}\and\ref{inst2}}
          \and
          C. Ramos Almeida\inst{\ref{inst1}\and\ref{inst2}}
          \and
          T. Ruiz-Lara\inst{\ref{inst1}\and\ref{inst2}}
          \and
          J. Falc\'on-Barroso\inst{\ref{inst1}\and\ref{inst2}}
          \and
          S.F. S\'anchez\inst{\ref{inst3}}
          \and
          P. S\'anchez-Bl\'azquez\inst{\ref{inst4}\and\ref{inst5}}
          \and
          I. M\'arquez\inst{\ref{inst6}}
          \and
          J. Masegosa\inst{\ref{inst6}}
          }

    \institute{Instituto de Astrof\'isica de Canarias, C/ V\'ia L\'actea, s/n, 38205, La Laguna, Tenerife, Spain \label{inst1}
                \and
                Departamento de Astrof\'isica, Universidad de La Laguna, 38206, La Laguna, Tenerife, Spain\label{inst2}
                \and
                Instituto de Astronom\'ia, Universidad Nacional Aut\'onoma de M\'exico, A. P. 70-264, C.P. 04510, CDMX, M\'exico\label{inst3}
                \and
                Departamento de F\'isica de la Tierra y Astrof\'isica, Universidad Complutense de Madrid, 28040, Madrid, Spain\label{inst4}
                \and
                IPARCOS, Facultad de CC F\'isicas, Universidad Complutense de Madrid, 28040, Madrid, Spain\label{inst5}
                \and
                Instituto de Astrof\'isica de Andaluc\'ia (CSIC), Glorieta de la astronom\'ia S/N, 18008 Granada, Spain\label{inst6}
                }

   \date{Accepted XXX. Received YYY; in original form ZZZ}

 
  \abstract

   \abstract{
    We present a comparison of the spin parameter $\lambda_R$, measured in a region dominated by the galaxy disc, between 20 pairs of nearby (0.005$<$z$<$0.03) seemingly isolated twin galaxies differing in nuclear activity. We find that 80--82\% of the active galaxies show higher values of $\lambda_R$ than their corresponding non-active twin(s), indicating larger rotational support in the AGN discs. This result is driven by the 11 pairs of unbarred galaxies, for which 100\% of the AGN show larger $\lambda_R$ than their twins. These results can be explained by a more efficient angular momentum transfer from the inflowing gas to the disc baryonic matter in the case of the active galaxies. This gas inflow could have been induced by disc or bar instabilities, although we cannot rule out minor mergers if these are prevalent in our active galaxies. This result represents the first evidence of galaxy-scale differences between the dynamics of active and non-active isolated spiral galaxies of intermediate stellar masses (10$^{10}<M_*<10^{11}$ M$_{\sun}$) in the Local Universe.} 


   \maketitle
%

\section{Introduction}\label{sec:Introduction}

\defcitealias{2019MNRAS.485.3794D}{dMC19}

Observational evidence suggests a co-evolution of the central supermassive black holes (SMBHs; see \citealt{Ho_2008} for a review) and their host galaxies (see e.g. \citealt{Kormendy_2013}). Unveiling the mechanism(s) controlling this co-evolution is crucial to improve our understanding of the formation and evolution of galaxies. Active galactic nuclei (AGN) feedback has been proposed as the main mechanism regulating SMBH and galaxy growth in massive galaxies, and it can act in different ways (e.g. \citealt{2012ARA&A..50..455F}). Understanding how nuclear activity is triggered and whether all massive galaxies go through an active phase is then of great importance.  
Significant observational effort has been made to observe the closest environment of SMBHs in order to study AGN fuelling (e.g. \citealt{2005A&A...441.1011G,Cris_2017,2019NatAs...3...48S} and references therein). To induce inflows of gas towards the galactic centre, the gas in the host galaxy has to lose angular momentum, but the dominant mechanism(s) that transport the gas from galaxy scales to the central parsecs in AGN are not clear yet 
\citep{2012NewAR..56...93A}. In this respect, the definition of non-active samples is essential to search for properties that might be unique to AGN. Legacy integral-field field spectroscopy (IFS) surveys such as ATLAS$^{3D}$ \citep{2011MNRAS.413..813C}, CALIFA \citep{2012A&A...538A...8S}, SAMI \citep{SAMI_2012}, MASSIVE \citep{MASSIVE_2014} or MaNGA \citep{MANGA_2015} now provide the opportunity of selecting almost identical pairs of galaxies differing only in nuclear activity. By comparing the kinematic properties of these pairs it should be possible to spot differences connected to AGN triggering \citep[hereafter dMC19]{2019MNRAS.485.3794D}.

An effective way of quantifying the global velocity structure of galaxies taking advantage of IFS is $\lambda_R$ \citep{Emsellem_2007}. This dimensionless spin parameter permits us to assess the rotational support of a galaxy, similarly to V/$\sigma$, and it goes to unity when rotation dominates.
Thanks to the previously mentioned IFS surveys it has been possible to study $\lambda_{R}$ for a large number of galaxies covering a wide range of  morphological types and stellar masses \citep{2016Cappellari, Colless_2018}. These data have been used to look for dependencies with different global galaxy properties. For example, \citet{Krajnovic_2013} studied a sample of early-type galaxies from ATLAS$^{3D}$ and found that $\lambda_{R}$ decreases with increasing bulge fraction. This dependence was confirmed for a wider range of galaxy morphologies by \citet{Falcon_2019} using data from CALIFA. 
An anti-correlation between $\lambda_{R}$ (or V/$\sigma$) and stellar mass (M$_*$) has also been reported using data from MASSIVE and ATLAS$^{3D}$ \citep{Veale_2017}, MaNGA \citep{Graham_2018} and SAMI \citep{van_de_Sande_2018}.
It is noteworthy that none of the latter works studied possible differences in $\lambda_R$ between active and non-active galaxies.
A few works based on large IFS surveys comparing different properties of active and non-active galaxies have been published (e.g. \citealt{2017MNRAS.472.4382R,Sebas_2018,2020MNRAS.492.3073L}). In \citet{Sebas_2018}, based on MaNGA data and including 98 AGN, the authors report that on average these AGN have lower rotational support ($\sim$65\%) than the non-active galaxies within 1R$_e$ (i.e. bulge-dominated). This trend is mainly driven by the late type galaxies, which dominate their AGN sample (see Fig. 2 in \citealt{Sebas_2018}). 
\citet{2019MNRAS.484..252I} performed a kinematic analysis of 62 AGN from MaNGA and reported larger differences between  $\sigma_{\rm gas}$ and $\sigma_{\rm stars}$ in the central kpc of the active galaxies than in the corresponding control sample. 

In \citetalias{2019MNRAS.485.3794D} we performed a pilot study based on two low-luminosity isolated active galaxies (barred and unbarred) and their two non-active twins matched in galaxy properties using data from the CALIFA survey. 
Our goal was to identify large-scale differences between the twins that could be related to AGN triggering. From the analysis of this pilot sample we found that each active galaxy had larger $\lambda_{R}$ within 1R$_e$ and globally, internal twists in their gas discs, dynamical lopsidedness and older stellar populations in the central kpc of the galaxy than its corresponding non-active twin. In view of these results, we decided to extend this study to a larger sample of twin galaxies differing in nuclear activity selected from CALIFA. This is the first work focused on studying the differences in stellar $\lambda_R$ between active and non-active galaxies matched in galaxy properties and based on one-to-one comparisons. In the following we assume a standard $\Lambda$CDM model with H$_{0}$ = 71\,km\,s$^{-1}$\,Mpc$^{-1}$, $\Omega_{m}$ = 0.27 and $\Omega_{\Omega}$ = 0.73.

\section{Observations and sample selection}\label{sec:Sample}

We use data from the third Data Release \citep{2016A&A...594A..36S} of the 
CALIFA survey \citep{2012A&A...538A...8S}, that corresponds to 667 galaxies in the Local Universe (0.005$<$z$<$0.03). \citet{2017A&A...598A..32M} selected 404 of these galaxies to characterise their 2D morphology discarding interacting systems, mergers and highly inclined galaxies (i$>$70\degr). This is the parent sample of this work. We use the fully reduced COMBO data cubes, which are a combination of those obtained with the V500 (wavelength range 3745-7500\,\AA, R$\sim$850) and V1200 (3700-4840\,\AA, R$\sim$1650) gratings. For two of the galaxies we do not have COMBO data cubes and we use the V500 instead. Details on the observational strategy, data quality, data reduction and statistical properties of the CALIFA survey can be found in \citet{2012A&A...538A...8S,2016A&A...594A..36S,2014A&A...569A...1W} and \citet{2015A&A...576A.135G}.

In order to identify isolated galaxies in CALIFA we followed the criteria of \citet{2014A&A...568A..70B}. We discarded galaxies having companions with similar systemic velocity (absolute difference smaller than 1000 km\,s$^{-1}$) and SDSS r-band magnitudes (within 2 mag) within a physical radius of 250\,kpc. By doing this we discard major mergers/interactions but we cannot rule out the presence of minor mergers.

To select our AGN sample we 
first used the codes \textsc{pPXF} \citep{2004PASP..116..138C, 2017MNRAS.466..798C} and \textsc{GANDALF} \citep{2006MNRAS.366.1151S, 2006MNRAS.369..529F} to characterize the emission line profiles of the central spaxel of each galaxy. We consider AGN candidates all the objects above the \citet{2001ApJ...556..121K} demarcation curve in the three classical BPT diagrams \citep{1981PASP...93....5B} that lie in the Seyfert region of the [OIII]/[OII] versus [NII]/H$\alpha$ diagram \citep{2010MNRAS.403.1036C}. Using these criteria we identified 19  isolated active galaxies with Hubble types from Sa/SBa to Sbc/SBbc. Here we focus on spiral galaxies because we are interested in the study of $\lambda_R$ of the stars in the disc component (see Section \ref{sec:Methods} for further details). These galaxies are low-luminosity AGN, making them ideal candidates to look for triggering imprints and characterise the stellar populations of their host galaxies.

For each AGN we then selected a control sample of non-active galaxies with similar galaxy properties (twin galaxies). By doing this any possible difference between the twins should be associated with nuclear activity. We matched the twins in Hubble type, stellar mass (M$_*$), absolute magnitude in the r-band (M$_{\rm R}$) and disc ellipticity in the r-band ($\epsilon$). Absolute differences are smaller than 0.25\,log(\,M$_*/M_{\sun}$), 0.70\,mag and 0.20, respectively (see Table \ref{tab:Sample}). 
Applying these criteria we did not find any twin for five of the 19 active galaxies. In the case of the barred active galaxies we selected twins with bar radius differences below 3\,kpc. With this restriction we have another three AGN lacking of twins. This leaves us with a final sample of 11 active spiral galaxies (ten type-2 and one type-1 AGN), of which 5 are barred and 6 unbarred.  

\begin{figure}
\begin{center} 

\includegraphics[width=0.45\textwidth]{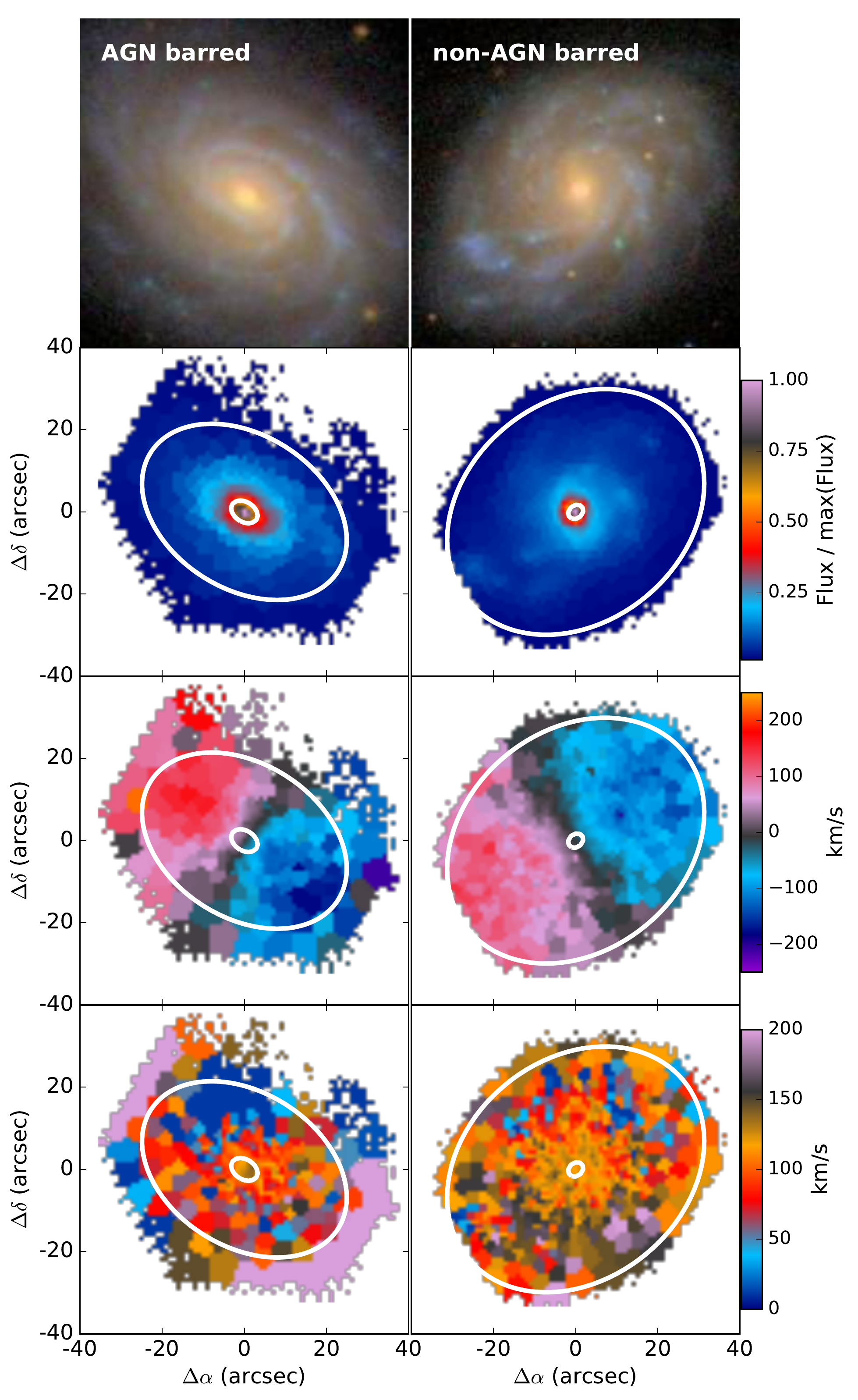}

\caption[]{Colour-composite SDSS images and CALIFA stellar flux, velocity and velocity dispersion maps (from top to bottom) of one pair of twin galaxies: NGC\,0214 (AGN; left column) and NGC\,2253 (non-AGN; right column). The white ellipses indicate the boundaries of disc region defined in Section \ref{sec:Methods} (2$R_e^{\rm bul} \leq R_{i}\leq  R_{\rm out}$).} \label{fig:pair}
 \end{center}
\end{figure}

Finally, we visually inspected the SDSS images of each active galaxy and its corresponding twin candidates to select the most similar one (best twin hereafter; see Fig. \ref{fig:pair} for an example) and discard only those that clearly have different appearances. After doing this we end up with five AGN with two or more non-active twins. In these cases, we always identified the best twin but we did not discard the others. Furthermore, five non-active galaxies are selected as twins of two different AGN. In total we have 20 pairs of isolated twin galaxies differing in nuclear activity (see Table \ref{tab:Sample} and Appendix \ref{app:Appendix_B}).

\begin{table*}
\begin{center}
\small
\begin{tabular}{ l c c c c c c c c c c c c}
\hline
\hline
AGN  & Type & D$_L$ & Scale & M$_*$  & M$_R$  & $\epsilon$   & r$_{\rm bar}$  & R$_e^{bul}$ &  R$_e$ & R$_{\rm out}$ & $\lambda_R$ & $\lambda_R^d$ \\
 \makebox[2.2cm][r]{Twin} &  & (Mpc) & (pc/\arcsec) & log(M/M$_{\odot}$) & (mag) &   & (kpc) & (kpc) & (kpc) & (kpc) & \\
 \makebox[1.5cm][r]{(1)}  &  (2) & (3) & (4) & (5)  & (6)  & (7)   & (8) &  (9) & (10) & (11) & (12) & (13) \\
\hline
NGC0214              &  SBbc  & 62  & 301 & 10.73 & -21.93 & 0.31 & 5.4    & 0.5  & 5.5 & 8.1 & 0.72 & 0.76$\pm$0.02  \\ 
 \makebox[2.5cm][r]{NGC2253} &  SBbc  & 53  & 257 & 10.50 & -21.34 & 0.21 & 3.9    & 0.3  & 4.0 & 8.7 & 0.53 & 0.61$\pm$0.03\\ 
NGC1093                &  SBbc  & 72  & 349 & 10.43 & -21.29 & 0.34 & 6.6    & 1.8  & 4.8 & 10.1 & 0.64 & 0.68$\pm$0.02\\ 
\makebox[2.65cm][r]{NGC5947*}               &  SBbc  & 83  & 402 & 10.56 & -21.16 & 0.18 & 6.4    & 0.5  & 5.0 & 12.1 & 0.58 & 0.67$\pm$0.03\\ 
\makebox[2.5cm][r]{NGC6004}               &  SBbc  & 62  & 301 & 10.63 & -21.41 & 0.18 & 9.0    & 0.3  & 6.7 & 9.4 & 0.60 & 0.69$\pm$0.03\\ 
\makebox[2.5cm][r]{NGC2253}               &  SBbc  & 53  & 257 & 10.50 & -21.34 & 0.21 & 3.9    & 0.3  & 4.0 & 8.7 & 0.53 & 0.61$\pm$0.03\\ 
\makebox[2.5cm][r]{NGC2540}                &  SBbc  & 91  & 441 & 10.21 & -21.49 & 0.38 & 7.1    & 0.5  & 6.3 & 12.6 & 0.67 & 0.71$\pm$0.02\\ 
NGC2410               &  SBb   & 68  & 330 & 10.86 & -21.60 & 0.72 & 6.3    & 1.1  & 7.1 & 13.0 & 0.76 & 0.76$\pm$0.01\\ 
\makebox[2.5cm][r]{NGC5522}                &  SBb   & 73  & 354 & 10.67 & -21.33 & 0.65 & 6.7    & 0.5  & 6.0 & 12.7 & 0.75 & 0.76$\pm$0.01\\ 
NGC2639               &  Sa    & 51  & 247 & 11.09 & -21.93 & 0.50 & --     & 1.0  & 4.3 & 8.4 & 0.62 & 0.65$\pm$0.01\\ 
\makebox[2.5cm][r]{NGC0160}                &  Sa    & 72  & 349 & 10.99 & -21.86 & 0.49 &  --    & 2.1  & 7.7 & 11.5 & 0.57 & 0.60$\pm$0.01\\ 
NGC2906                &  Sbc   & 34  & 165 & 10.46 & -20.60 & 0.44 & --     & 0.3  & 3.2 & 5.9 & 0.73 & 0.75$\pm$0.01\\ 
\makebox[2.65cm][r]{NGC0001*}                &  Sbc   & 63  & 305 & 10.58 & -21.30 & 0.38 & --     & 1.4  & 3.9 & 10.1 & 0.57 & 0.61$\pm$0.02\\ 
\makebox[2.5cm][r]{NGC6063}                 &  Sbc   & 48  & 233 & 10.28 & -20.37 & 0.45 &  --    & 0.9   & 4.8 & 7.9 & 0.70 & 0.72$\pm$0.01\\ 
\makebox[2.6cm][r]{UGC09777}                &  Sbc   & 75  & 364 & 10.25 & -20.60 & 0.43 & --     & 1.0  & 3.6 & 8.4 & 0.55 & 0.58$\pm$0.02\\ 
NGC2916                &  Sbc   & 57  & 276 & 10.64 & -21.25 & 0.35 &  --    & 0.9  & 7.2 & 10.1 & 0.71 & 0.75$\pm$0.02\\ 
\makebox[2.65cm][r]{NGC0001*}                &  Sbc   & 63  & 305 & 10.58 & -21.30 & 0.38 & --     & 1.4  & 3.9 & 10.1 & 0.57 & 0.61$\pm$0.02\\ 
NGC6394                &  SBbc  & 123 & 596 & 10.86 & -21.79 & 0.59 & 17.9   & 0.6  & 8.7 & 17.0 & 0.82 & 0.83$\pm$0.01\\ 
\makebox[2.60cm][r]{UGC12810}               &  SBbc  & 112 & 543 & 10.81 & -21.76 & 0.61 & 19.6   & 0.6  & 11.2 & 16.3 & 0.80 & 0.81$\pm$0.01\\ 
NGC7311                   &  Sa    & 64  & 310 & 10.96 & -22.20 & 0.48 &  --    & 0.7  & 3.9 & 9.9 & 0.70 & 0.72$\pm$0.01\\ 
\makebox[2.5cm][r]{NGC0160}                &  Sa    & 72  & 349 & 10.99 & -21.86 & 0.49 &  --    & 2.1  & 7.7 & 11.5 & 0.57 & 0.60$\pm$0.01\\ 
NGC7466                 &  Sbc   & 105 & 509 & 10.68 & -21.69 & 0.61 & --     & 2.0  & 6.9 & 16.3 & 0.78 & 0.79$\pm$0.01\\ 
\makebox[2.65cm][r]{NGC2596*}               &  Sbc   & 87  & 422 & 10.87 & -21.44 & 0.65 &  --    & 2.8  & 8.2 & 13.5 & 0.73 & 0.74$\pm$0.01\\ 
\makebox[2.5cm][r]{NGC5980}                &  Sbc   & 66  & 320 & 10.69 & -21.70 & 0.63 &  --    & 1.1  & 5.6 & 10.2 & 0.74 & 0.75$\pm$0.01\\ 
UGC00005                &  Sbc   & 100 & 485 & 10.74 & -21.90 & 0.49 & --     & 0.6  & 8.1 & 16.2 & 0.83 & 0.84$\pm$0.01\\ 
\makebox[2.65cm][r]{NGC2596*}               &  Sbc   & 87  & 422 & 10.87 & -21.44 & 0.65 &  --    & 2.8  & 8.2 & 13.5 & 0.73 & 0.74$\pm$0.01\\ 
\makebox[2.5cm][r]{NGC5980}               &  Sbc   & 66  & 320 & 10.69 & -21.70 & 0.63 &  --    & 1.1  & 5.6 & 10.2 & 0.74 & 0.75$\pm$0.01\\ 
\makebox[2.65cm][r]{NGC0001*}                &  Sbc   & 63  & 305 & 10.58 & -21.30 & 0.38 & --     & 1.4  & 3.9 & 10.1 & 0.57 & 0.61$\pm$0.02\\ 
UGC03973 $^{\dagger}$ &  SBbc  & 95  & 461 & 10.21 & -21.61 & 0.16 & 16.6   & 0.8  & 7.9 & 13.8 & 0.45 & 0.56$\pm$0.04\\ 
\makebox[2.75cm][r]{UGC02311*}             &  SBbc  & 97  & 470 & 10.43 & -21.85 & 0.33 & 13.6   & 0.7  & 6.7 & 12.1 & 0.63 & 0.68$\pm$0.02\\ 
\makebox[2.5cm][r]{NGC6032}                &  SBbc  & 69  & 335 & 10.42 & -20.96 & 0.26 & 16.4   & 0.4  & 9.2 & 11.9 & 0.67 & 0.72$\pm$0.02\\ 
\hline
\end{tabular}

\caption[Sample from CALIFA]{Sample properties. (1) Galaxy name (non-active twin names are shifted). * Indicates the best twin of each AGN and $^{\dagger}$ the only type 1 AGN in our sample. (2) Hubble type from \citet{2014A&A...569A...1W}, (3) luminosity distance and (4) spatial scale, (5) total stellar mass and (6) Petrosian magnitude from \citet{2014A&A...569A...1W}, (7) disc ellipticity, (8) bar radius and (9) bulge effective radius in the r-band from \citet{2017A&A...598A..32M}, (10) effective radius in the r-band from \cite{Falcon_2017}, (11) outer radius (this work), (12) projected $\lambda_R$ in the region dominated by the disc, with maximum associated errors of 0.01 and (13) deprojected $\lambda_R$ and corresponding error.  
}
\label{tab:Sample} 
\end{center}
\end{table*}

\section{Methodology}\label{sec:Methods}

The stellar kinematics were calculated following the same methodology as in \citetalias{2019MNRAS.485.3794D}. We removed all the spaxels with S/N<3 and spatially binned the spaxels using the Voronoi 2D binning method \citep{2003MNRAS.342..345C} to achieve S/N of $\sim$30. We used pPXF to fit the 3800--7000\,\AA~spectral range. Finally, $\lambda_R$  was measured as in \citet{Emsellem_2007}:
\begin{eqnarray}
    \lambda_{R} \equiv \frac{\langle R | V | \rangle}{\langle R \sqrt{V^{2} + \sigma^{2}} \rangle} = \frac{\displaystyle\sum_{i=1}^{N_{p}} F_{i} R_{i} |V_{i}| }{\displaystyle\sum_{i=1}^{N_{p}} F_{i} R_{i} \sqrt{V_{i}^{2} + \sigma_{i}^{2}} }
\end{eqnarray}
where R$_{i}$, F$_{i}$, V$_{i}$ and $\sigma_{i}$ are the distance to the centre, the flux, the stellar velocity and the velocity dispersion per spatial bin i.

$\lambda_R$ is usually calculated within one effective radius or radially (see e.g. \citealt{Falcon_2019} and references therein). 
By comparing the stellar $\lambda_R$ radial profiles of the two AGN and their non-active twins, in \citetalias{2019MNRAS.485.3794D} we found the largest differences in the disc of the galaxies. Furthermore, the interpretation given in \citetalias{2019MNRAS.485.3794D} to explain the differences in $\lambda_R$ between twin galaxies differing in nuclear activity also suggested to focus on the region dominated by the disc (see Section \ref{sec:Results}). 
To identify this region in each galaxy we used the two-dimensional photometric decomposition from \citet{2017A&A...598A..32M}. This way we can determine the radius at which the intersection between the bulge and disc profiles happens. We find that this radius is never larger than 2 times the bulge effective radius (R$_e^{\rm bul}$). Therefore, we can select 2\,R$_e^{\rm bul}$ as the inner radius of the region where the disc component dominates over the bulge for all the galaxies in the sample. To determine the disc outer radius (R$_{\rm out}$) we have to consider the extent of the galaxies in the CALIFA data. We define R$_{\rm out}$ as the radius of the outermost ellipse without empty spaxels/voxels (see Table \ref{tab:Sample} and Fig. \ref{fig:pair}). Thus, we used the annular region between 2\,R$_e^{\rm bul}$ and R$_{\rm out}$ to calculate $\lambda_R$ in the disc\footnote{For NGC\,2540, \citet{2017A&A...598A..32M} did not include a bulge in their photometric decomposition, but a nuclear point source. We thus estimated R$_e^{\rm bul}$ using the average ratio between R$_{\rm out}$/R$_e^{\rm bul}$ of all the galaxies with the same morphological type (SBbc) in our sample.}.

In order to calculate uncertainties we used Monte-Carlo simulations adding the corresponding errors of the CALIFA data cubes to the spectrum of each voxel following a Gaussian distribution. The procedure was repeated 100 times, resulting in 100 different $\lambda_R$ values. The standard deviation of this distribution of values is our statistical error, which is always smaller than 0.01. In the following we will consider 0.01 as the error of each $\lambda_R$.

\section{Results and discussion}\label{sec:Results}

\begin{figure*}
\begin{center} 

\includegraphics[width=\textwidth]{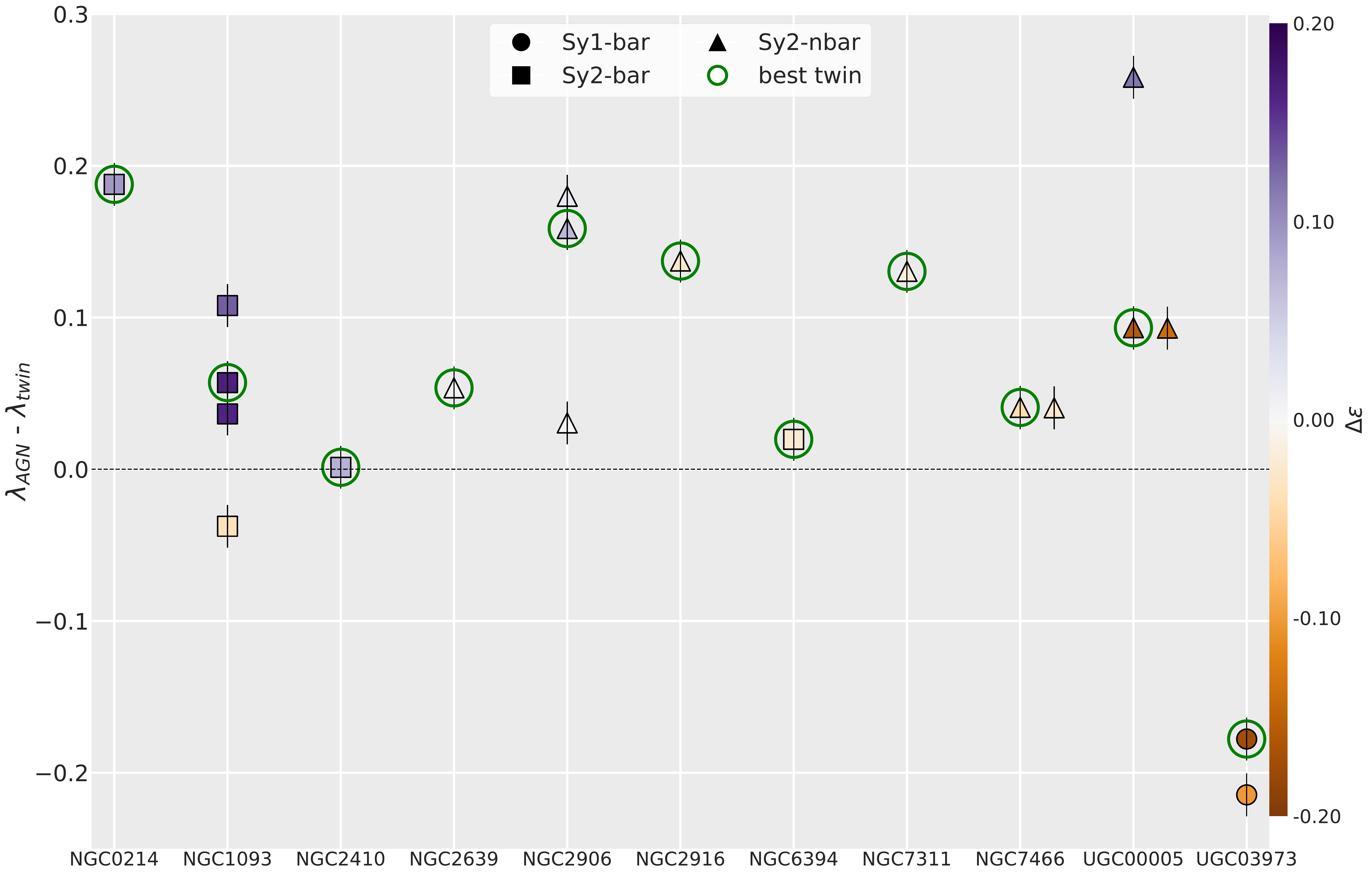}

\caption[]{Differences in stellar $\lambda_R$ between the pairs of twin galaxies. The best twin of each AGN is marked with a green circle. Each column corresponds to an active galaxy and each symbol to the difference in $\lambda_R$ with each of its twins. The colour code indicates the difference in ellipticity ($\epsilon_{AGN} - \epsilon_{twin}$, see Table \ref{tab:Sample}). Error bars correspond to propagation of the individual uncertainties (see Sect. \ref{sec:Methods}).} \label{fig:ang_mom_ellipticity}
 \end{center}
\end{figure*}

Figure \ref{fig:ang_mom_ellipticity} shows the differences in $\lambda_R$ of the disc ($\Delta\lambda_R$) between each AGN and its twin(s).
Considering the best twins only (indicated with green circles in  Fig. \ref{fig:ang_mom_ellipticity}), we find 9/11 (82\%) pairs in which the active galaxy has higher $\lambda_{R}$ than its twin. 
If we consider the 20 pairs of galaxies the percentage is similar: 16/20 (80\%) have $\Delta\lambda_R$>0.01 (i.e. larger than the corresponding error). The case of UGC\,03973, the only type 1 AGN in our sample, is noteworthy because it has considerably lower $\lambda_R$ than its two twins. This could be related with nuclear contamination of the central spaxels associated with the large PSF size of the CALIFA data. It is worth mentioning, however, that using a sample of 98 AGN from MaNGA, \citet{Sebas_2018} reported a tendency of the type 1 AGN to be hosted in more massive, centrally concentrated and pressure-supported galaxies (within 1R$_e$) than type 2 AGN. If these results are confirmed it would provide support to an evolutionary transition between types rather than mere orientation-based differences. A larger AGN sample is required to investigate whether type 1s have different kinematics than type 2s. 

As we mentioned in Section \ref{sec:Sample}, we have five barred AGN and six without bars. 
If we only consider the unbarred galaxies in our sample, we find that 11/11 pairs (100\%) show $\Delta\lambda_R$>0.01. In the case of the barred galaxies, we have 5/9 pairs (56\%), 5/7 (71\%) if we do not consider the type 1 AGN. 
 
$\lambda_R$ is derived from projected quantities and therefore it depends on the viewing angle \citep{Emsellem_2007}. However, for our twins the effect of galaxy inclination on $\Delta\lambda_{R}$ should be small because we matched them in disc ellipticity ($\epsilon$). In Fig. \ref{fig:ang_mom_ellipticity} we show with different colours the differences in ellipticity ($\Delta\epsilon$) between each AGN and its corresponding twin(s), which are always smaller than 0.2 (see Section \ref{sec:Sample}). Seven active galaxies have higher $\epsilon$ than their twins (i.e. more inclined; purple symbols), seven have lower $\epsilon$ (brown symbols) and four are almost identical (white symbols). Thus, the slightly different inclinations between twins would not be driving the difference in $\lambda_R$. To furher confirm this, we deprojected the
individual $\lambda_R$ values as described in Appendix B. Considering these values and their corresponding errors (see Table \ref{tab:Sample}), the percentage of AGN having larger $\lambda_R$ than their best twins remains the same (82\%) and it is 70\% when we consider all the pairs (14/20). Thus we confirm that the effect of inclination on $\Delta\lambda_R$ for the galaxies in our sample is small. It is noteworthy that the six pairs in which the AGN have smaller or equal $\lambda_{R}$ than its twin(s) are all barred. For the unbarred twins, we continue having positive $\Delta\lambda_{R}$ in 100\% of the pairs after deprojecting.

The stellar ages of the discs could also have an influence on $\lambda_R$, since there is a relation between age and V/$\sigma$ (the older the stellar population the lower V/$\sigma$ and viceversa; e.g. \citealt{van_de_Sande_2018}).  
In order to explore this possibility we characterised the stellar populations of the galaxies in the same region as $\lambda_R$ using the code \textsc{STECKMAP} \citep{2006MNRAS.365...46O, 2006MNRAS.365...74O}. We used a similar methodology as in \citetalias{2019MNRAS.485.3794D}.
In Appendix \ref{app:Appendix} we show an alternative version of Fig. \ref{fig:ang_mom_ellipticity} indicating with different colours the differences in light-weighted averaged stellar age. In general, the discs of AGN are older than those of their twins (13/20; red symbols in Fig. \ref{fig:ang_mom_age}; see also \citealt{2020MNRAS.492.3073L}), whilst the other seven show equal or younger ages (white and blue symbols respectively). Maximum differences are $\pm$1.4 Gyr. This is the opposite of what we would expect if the differences in $\lambda_R$ would be driven by the differences in the stellar ages of their discs. We also note that for some AGN with more than one twin, the differences in stellar age are both positive and negative (e.g. NGC\,1093, UGC\,00005 and UGC\,03973).  This constitutes further indication that differences in stellar age are not driving our result. A detailed study of the stellar populations of these galaxies including the central regions will be the subject of a forthcoming work (del Moral-Castro in prep.).

Another parameter that could have an influence on $\lambda_{R}$ is the bulge fraction \citep{Krajnovic_2013}, with the larger the bulge to total flux ratios (B/T) the lower the $\lambda_R$. Although our twins are matched in Hubble type and M$_*$ and here we compare $\lambda_{R}$ in the region dominated by the disc, we evaluate whether possible differences in the B/T fractions might have any influence in our result. In Appendix \ref{app:Appendix} we show another version of Fig. \ref{fig:ang_mom_ellipticity} with different colours indicating the differences in B/T measured from SDSS r-band photometry \citep{2017A&A...598A..32M}. Nine active galaxies have higher B/T than their corresponding twins and ten have lower values\footnote{We note that NGC\,2540 does not have B/T value (see Section \ref{sec:Methods}).} (see Fig. \ref{fig:ang_mom_BT}). Comparing galaxies with relatively large differences in effective radius (R$_{e}$) could also have an impact on $\Delta\lambda_R$. There are eight AGN with higher R$_{e}$ than their twins and twelve showing lower R$_{e}$. Considering only the 13 pairs with smaller R$_{e}$ differences than 30\% of 
the AGN R$_{e}$, we have 9/13 pairs (69\%) having $\Delta\lambda_R$>0.1 and 6/8 pairs (75\%) if we only consider the best twins.
Finally, galaxies with larger M$_*$ generally have lower $\lambda_R$ values (e.g. \citealt{Veale_2017}). In our case, 10 AGN have larger M$_*$ than their twins, and 10 have lower M$_*$ (see Table \ref{tab:Sample}). Thus, the differences in stellar $\lambda_R$ are not driven by smaller bulge fraction, R$_e$ or M$_*$ of the AGN relative to their twins.

The results here presented provide statistical support to the preliminary interpretation of our pilot study (\citetalias{2019MNRAS.485.3794D}). We identify the higher $\lambda_R$ measured in the discs of isolated AGN as the imprint of the angular momentum transfer from the inflowing gas to the baryonic matter in the disc \citep{Kormendy_winter_2013, Saha_2014}.  
A large-scale disc or bar instability would have induced gas to move from the disc to the central region, triggering nuclear activity. 
For this inflow to happen, the gas had to efficiently transfer angular momentum to the baryonic matter remaining in the disc to conserve the total angular momentum of the isolated system. As a consequence of this internal angular momentum redistribution, the galactic disc got dynamically colder (i.e. higher $\lambda_R$). In the case of the barred galaxies, both active and non-active, the bars promote the inflow of gas, transporting angular momentum outwards (e.g. \citealt{Kormendy_winter_2013}). 
This makes it more challenging to detect differences in $\lambda_R$ between twins than in the case of the non-barred
galaxies. 

Another possibility is that the gas has an external origin (i.e. minor mergers) in the case of the active galaxies. 
As we mentioned in Section \ref{sec:Sample}, our sample does not include major mergers and clearly interacting systems, but based on the SDSS images only we cannot rule out the presence of minor mergers (see \citealt{2010MNRAS.408..400S} for a comparison between SDSS and deep optical images of apparently isolated Seyfert galaxies). Thus, gas-rich minor mergers constitute another possible explanation for the results presented here if they are prevalent in AGN: not only they promote spiral structure \citep{2011Natur.477..301P} therefore increasing the disc $\lambda_R$, but also provide gas supply that can potentially trigger nuclear activity (e.g. \citealt{2014MNRAS.437.3373N,2014MNRAS.445L..51T}). 
Deeper optical imaging is required to detect low surface brightness features indicative of merging activity \citep{2011MNRAS.410.1550R}. 

In the framework of these interpretations, it would be also relevant to look for differences in the actual specific angular momenta of the active/non-active twins and not only in $\lambda_R$.
To do that we use j$_*$, the stellar specific angular momentum (\citealt{1983IAUS..100..391F,Romanowsky_2012}; see Appendix \ref{app:Appendix_j_p}). To compare with $\lambda_R$, we computed j$_*$ in the disc dominated region and show the j$_*$--M$_*$ diagram in Appendix \ref{app:Appendix_j_p}. We find that 50\% of the pairs show positive/negative differences in j$_*$, and this is also valid if we calculate the total j$_*$ values (i.e, including the bulge-dominated region). This would imply that AGN are triggered in discs where a redistribution of angular momentum has happened (resulting in similar j$_*$ and higher $\lambda_R$) but not necessarily in galaxies with higher/lower angular momentum or with higher/lower circular velocities (and thus more/less massive haloes).
Nevertheless, we cannot rule out an underestimation of the j$_*$ values calculated for our sample due to the limited field-of-view of the CALIFA data. According to \citet{Romanowsky_2012} at least 2\,R$_{e}$ are necessary to estimate j$_*$ reliably\footnote{This is not the case for $\lambda_R$, which is usually measured within 1\,R$_{e}$.}, and 16 out of the 25 galaxies in our sample have R$_{\rm out}$<2\,R$_{e}$.

Finding galaxy-scale differences between the active galaxies and their non-active twins appears puzzling because AGN are now understood as a short and likely episodic phase of galaxy evolution. AGN lifetimes are estimated to be $\leq$100 Myr (e.g. \citealt{2004cbhg.symp..169M,2005ApJ...630..716H}), which represents a tiny fraction of the time that any change in galaxy morphology and dynamics might take (e.g. \citealt{2005SciAm.293d..42C,2008MNRAS.391.1137L}). Therefore, if all SMBHs go through an active phase we should not expect any large-scale difference between the twins. The result shown in Fig. \ref{fig:ang_mom_ellipticity} could then imply that not every galaxy goes through an active fase, at least in the redshift and mass range considered here.

This letter shows, for the first time, tentative evidence that the discs of AGN in seemingly isolated spiral galaxies of intermediate stellar masses (10$^{10}<M_*<10^{11}$ M$_{\sun}$) present larger rotational support than their non-active twins. Performing one-to-one comparisons rather than the commonly used AGN vs. control sample studies was fundamental to spot this. 
This result needs to be further explored and confirmed for a larger sample of active and non-active galaxies, preferably using integral field data of higher angular and spectral resolution and larger spatial coverage. 

\begin{acknowledgements}
We thank R. Morganti, C. Tadhunter, F. Combes, C. Lagos, J. M\'endez-Abreu and V. Kalinova for useful suggestions. We thank the referee for carefully reading our manuscript and providing constructive comments.
This study uses data provided by the Calar Alto Legacy Integral Field Area (CALIFA) survey (\url{http://califa.caha.es/}). Based on observations collected at the Centro Astron\'omico Hispano Alem\'an (CAHA) at Calar Alto, operated jointly by the MPIA and the IAA-CSIC. 
IMC acknowledges the support of the IAC.
BGL, CRA, TRL, PSB, IM and JM acknowledge support from the Spanish Ministry of Science, Innovation and Universities (MCIU), the Agencia Estatal de Investigaci\'on (AEI) and the Fondo Europeo de Desarrollo Regional (EU FEDER) under projects with references PID2019-107010GB-I00, PID2019-106027GB-C42, AYA2015-68217-P, AYA2016-76682-C3-2-P, AYA2017-89076-P, AYA2016-77237-C3-1-P, AYA2015-63810-P and AYA2016-76682-C3-1-P. CRA and TRL also acknowledge support from the MCIU under grants with reference RYC-2014-15779 and FJCI-2016-30342, respectively. 
IMC, BGL, CRA, TRL, IM and JM acknowledge financial support from the State Agency for Research of the Spanish MCIU through the "Center of Excellence Severo Ochoa" award to the Instituto de Astrofísica de Canarias (SEV-2015-0548) and to the Instituto de Astrofísica de Andalucía (SEV-2017-0709). SFS thanks CONACYT CB-285080 and FC-2016-01-1916, and PAPIIT-IN100519 projects for their support.

\end{acknowledgements}

%
   \bibliographystyle{bibtex/aa} 
   \bibliography{biblio} 
%

\begin{appendix} 

\section{SDSS images}\label{app:Appendix_B}

Here we include the colour-composite SDSS images of our sample of active and non-active twin galaxies  (Figs. \ref{fig:galaxy_images_1} and \ref{fig:galaxy_images_2}).

\begin{figure*}
\begin{center}

\includegraphics[width=\textwidth]{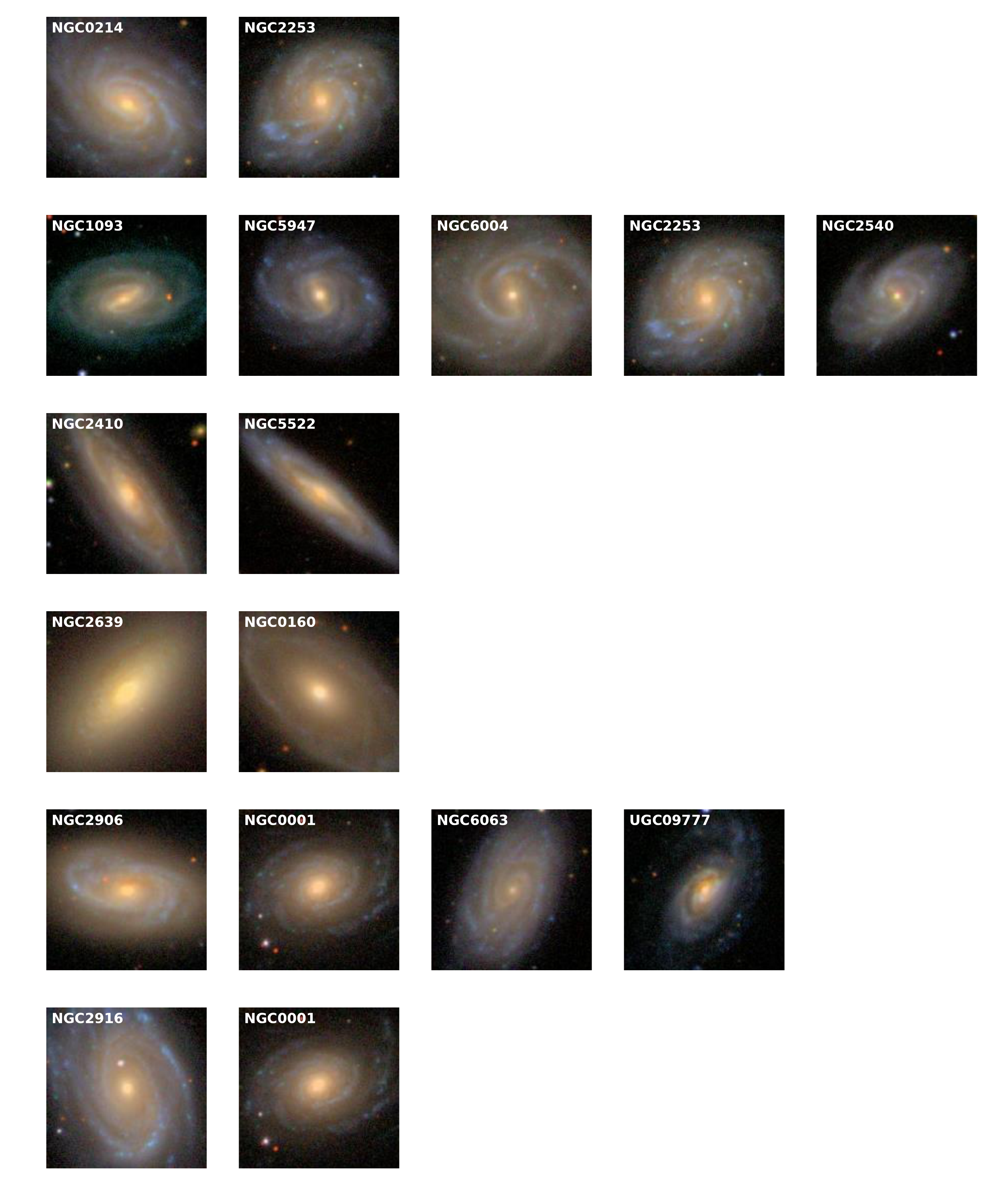}

\caption[]{Colour-composite SDSS images of the active galaxies (left column) and their corresponding non-active twin(s). For AGN with more than one twin, the best twin is the one shown next to the AGN. Each image has a field-of-view of 90\arcsec$\times$90\arcsec. North is up and East to the left.}

\label{fig:galaxy_images_1}
 \end{center}
\end{figure*}

\begin{figure*}
\begin{center} 

\includegraphics[width=\textwidth]{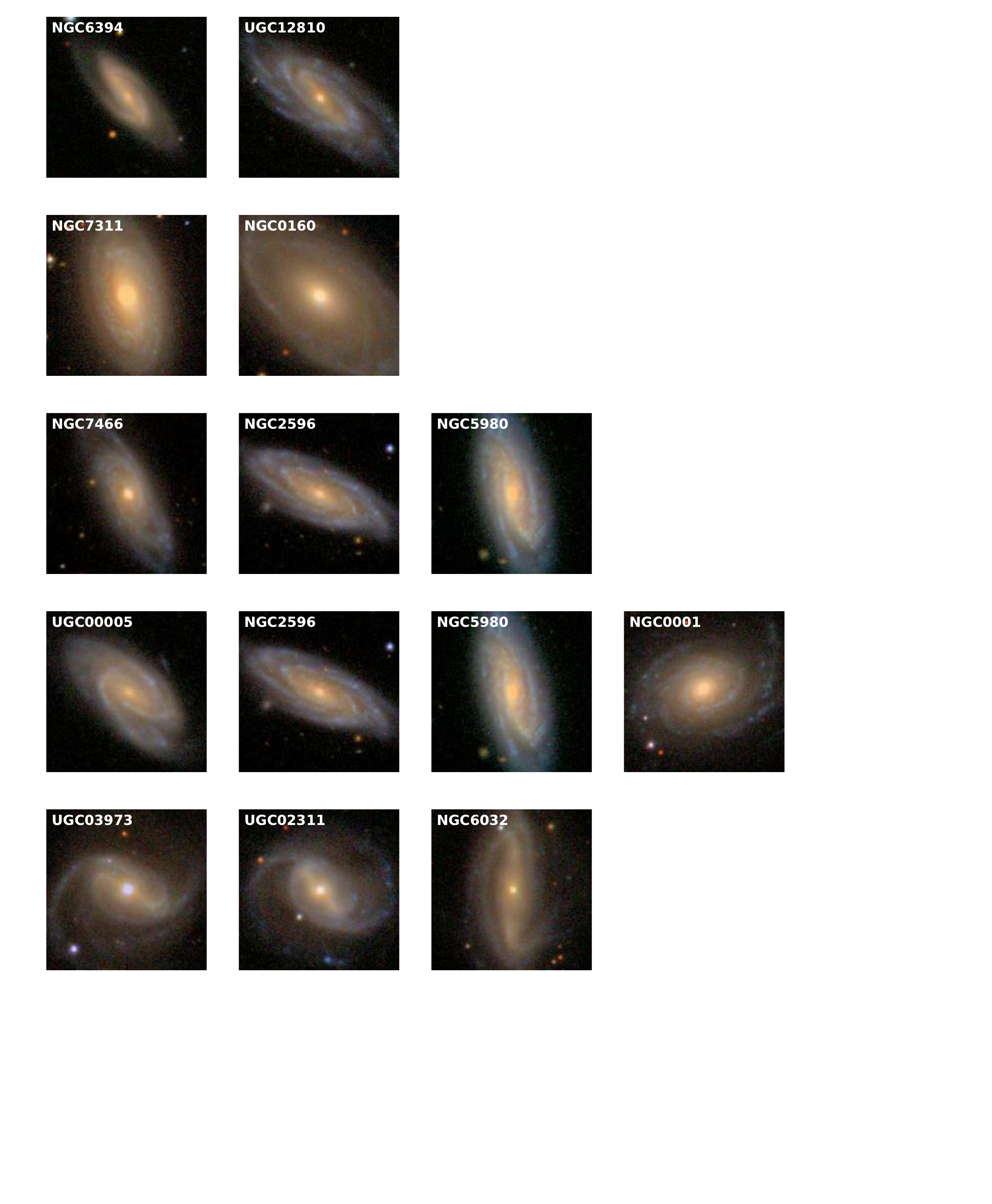}

\caption[]{Continuation of Fig. \ref{fig:galaxy_images_1}.}

\label{fig:galaxy_images_2}
 \end{center}
\end{figure*}

\section{Deprojected $\lambda_{R}$ values}\label{app:Appendix_deprojected}

To deproject the individual $\lambda_{R}$ values we use the following equation from Appendix B of \citet{Emsellem_2011}
\begin{eqnarray}
    \lambda_{R}^d = \frac{\lambda_{R}}{\sqrt{C^2 - \lambda^{2}_{R} (C^{2} - 1)}} 
\end{eqnarray}
where $C = \sin{i}/\sqrt{1 - \beta \cos^{2}{i}}$, i the galaxy inclination and $\beta$ the anisotropy parameter. To estimate C for each galaxy we used the median value of $\beta$ for its Hubble type ($\beta_{Sa}$=0.30, $\beta_{Sb}$=0.34, $\beta_{Sbc}$=0.29 from Table B1 in \citealt{Kalinova_2017}). The individual $\lambda_R^d$ are reported in Table\,\ref{tab:Sample}.  
This deprojection introduces additional errors, including assumptions on internal galaxy dynamics. In an attempt to quantify this, we have estimated conservative errors by taking into account the ellipticity errors from \citet{2017A&A...598A..32M}, the statistical $\lambda_R$ uncertainties and the $\beta$ standard deviation of the corresponding Hubble type. 

\section{Alternative figures}\label{app:Appendix}

\begin{figure*}
\begin{center} 

\includegraphics[width=.9\textwidth]{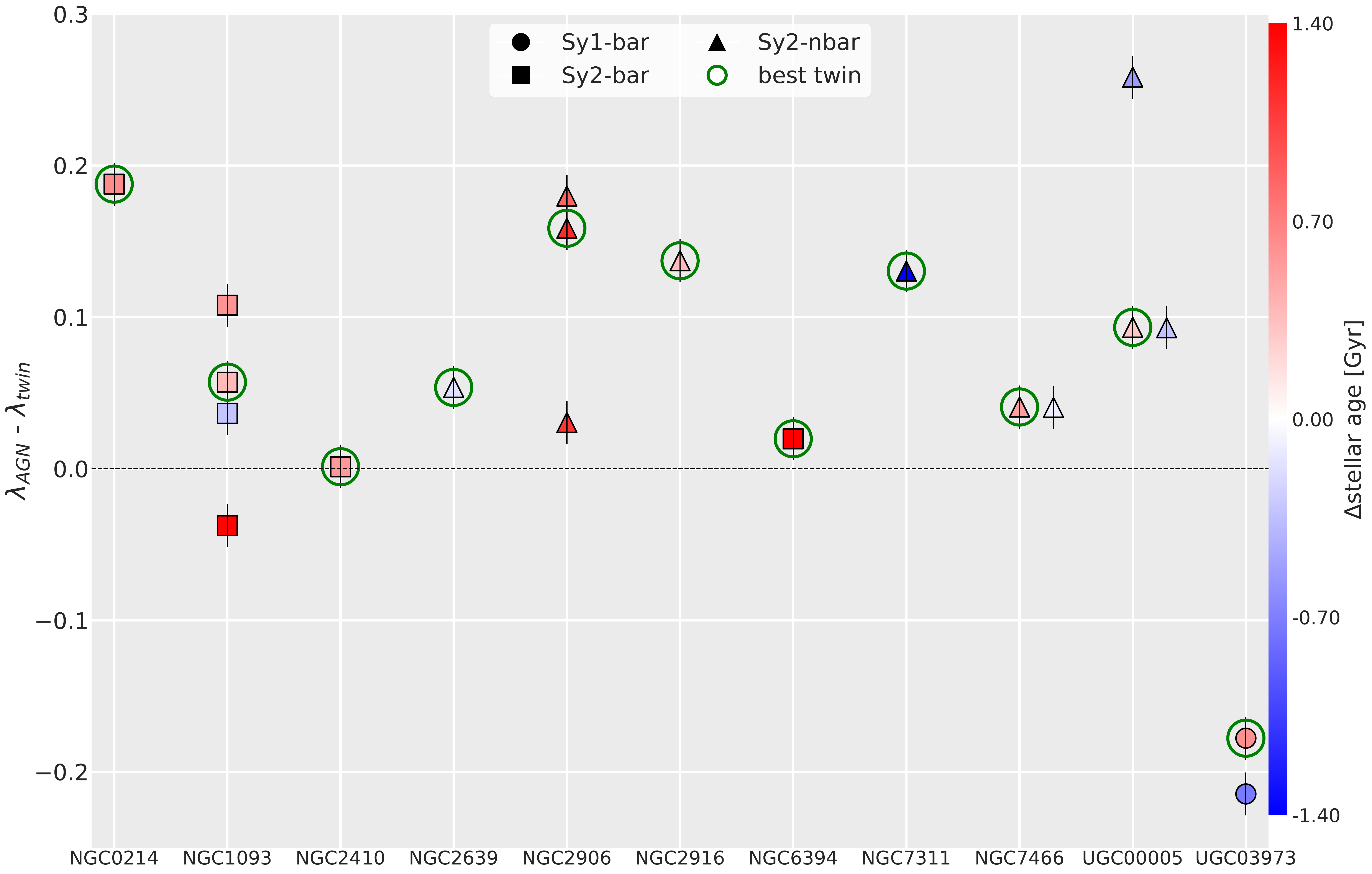}

\caption[]{Same as Fig. \ref{fig:ang_mom_ellipticity} but with the colour code indicating differences in averaged luminosity-weighted stellar age measured in the same region as $\lambda_R$.}
\label{fig:ang_mom_age}
 \end{center}
\end{figure*}

\begin{figure*}
\begin{center} 

\includegraphics[width=.9\textwidth]{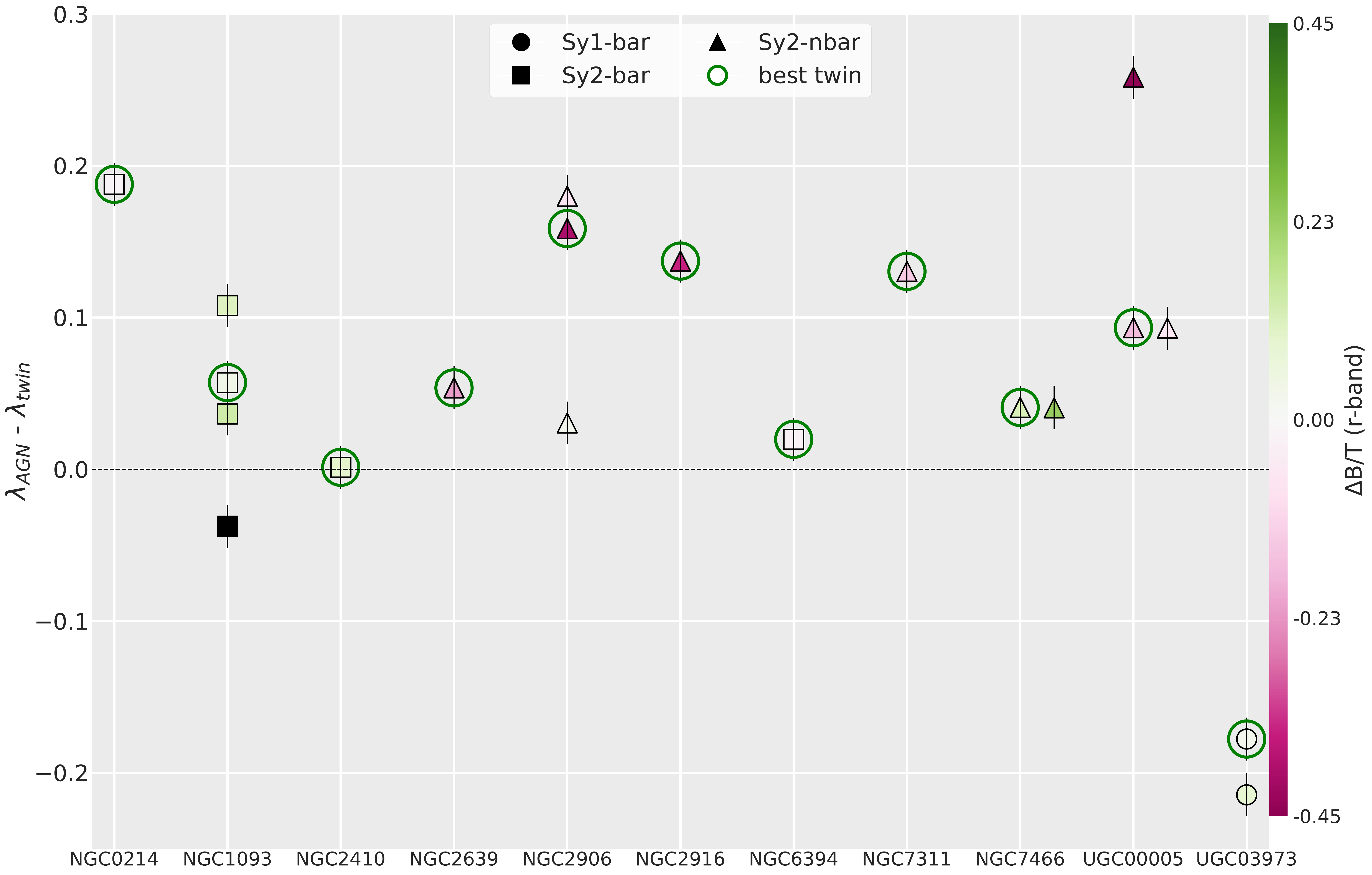}

\caption[]{Same as in Fig. \ref{fig:ang_mom_ellipticity} but with the colour code indicating differences in B/T from SDSS r-band photometry \citep{2017A&A...598A..32M}. The best twin of each AGN is indicated with a red circle. NGC\,2540 does not have B/T value reported in \citet{2017A&A...598A..32M} and thus its corresponding $\Delta\lambda_R$ appears as a black square.}
\label{fig:ang_mom_BT}
 \end{center}
\end{figure*}

Here we include two alternative versions of Fig.  \ref{fig:ang_mom_ellipticity}, with different colours indicating differences in light-weighted averaged stellar age (Fig.  \ref{fig:ang_mom_age}) and in B/T from SDSS r-band photometry (Fig. \ref{fig:ang_mom_BT}) between the twins.

\section{Total specific angular momentum}\label{app:Appendix_j_p}

To compute the j$_*$ values of our galaxies as a proxy for the specific angular momentum (J), we use the methodology outlined by \citet{Romanowsky_2012} and adapted by \citet{Cortese_2016} to IFS data as it follows
\begin{eqnarray}
    j_*  = \frac{\displaystyle\sum_{i=1}^{N} F_{i} R_{i} |V_{i}| }{\displaystyle\sum_{i=1}^{N} F_{i} }
\end{eqnarray}
where R$_{i}$, F$_{i}$ and V$_{i}$ are the galactocentric radius, flux and stellar velocity per spatial bin. We calculate the j$_*$ values in the same region than $\lambda_{R}$ (i.e. disc dominated; see Section \ref{sec:Methods}).  

\begin{figure*}
\begin{center} 

\includegraphics[width=\textwidth]{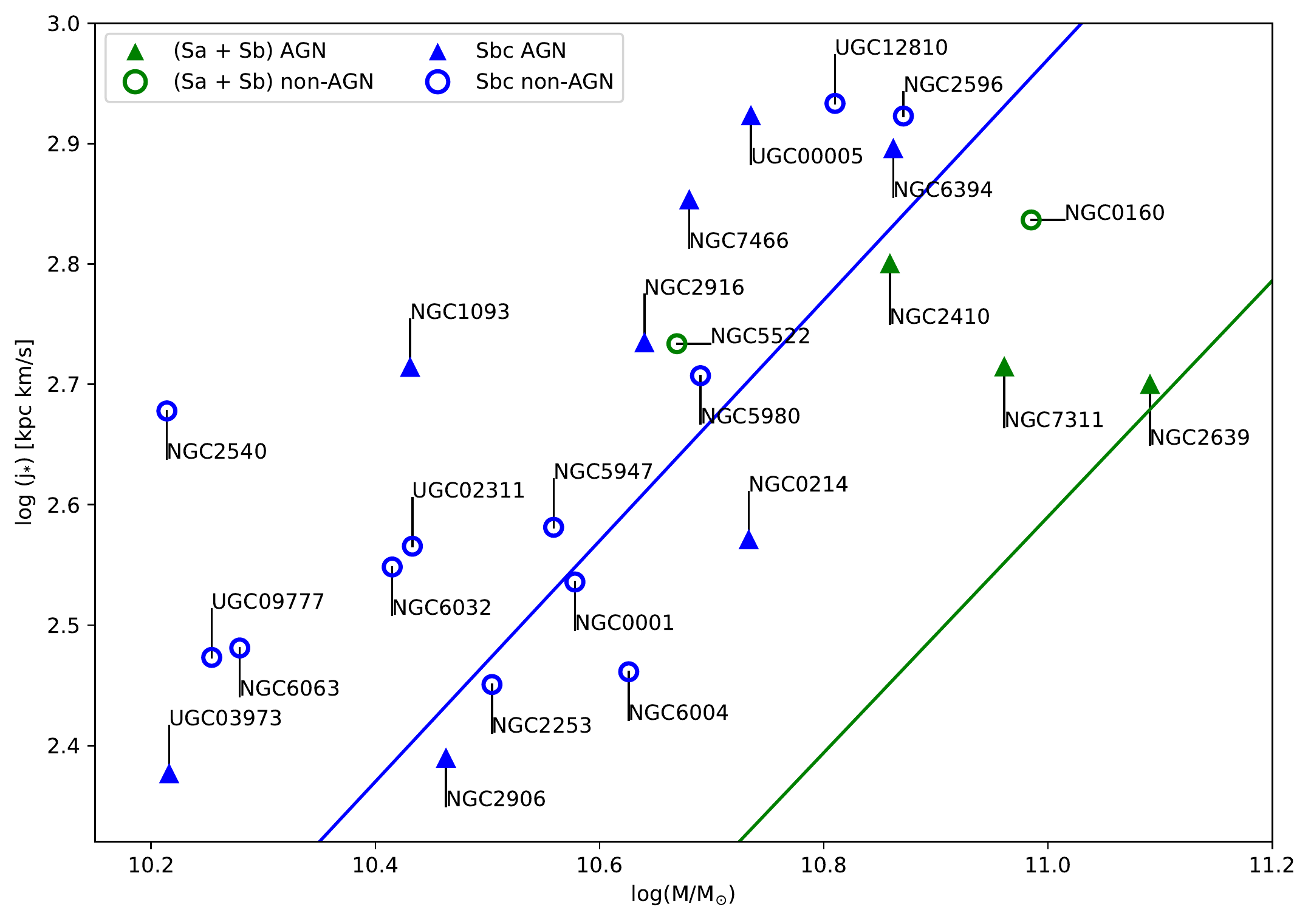}

\caption[]{j$_{*}$--M$_{*}$ diagram of the active and non-active galaxies in our sample. j$_{*}$ values corresponds to the disc dominated region defined in Section \ref{sec:Methods}. Filled symbols correspond to AGN and open symbols to non-active galaxies. Colours correspond to Hubble type (green to Sa and SBb and blue to S(B)bc galaxies). The green and blue lines correspond to the linear fits to S0/Sa-Sb and Sbc/later types from \citet{Cortese_2016} within 1R$_e$ for comparison.}
\label{fig:j_p_profiles}
 \end{center}
\end{figure*}

In Figure \ref{fig:j_p_profiles} we show the j$_*$--M$_*$ diagram of the active and non-active galaxies in our sample. The scatter of this plot is strongly correlated with galaxy morphology \citep{1983IAUS..100..391F,Cortese_2016}, something that we also see for our galaxies, which include Sa, SBb and S(B)bc types.

\begin{table}
\begin{center}
\small
\begin{tabular}{ l c }
\hline
\hline
AGN  & $\log$\,(j$_{*}$) \\
 \makebox[2.2cm][r]{Twin} &  (kpc km/s)  \\
 \makebox[1.5cm][r]{(1)}  &  (2)  \\
\hline
NGC0214                         & 2.57    \\ 
 \makebox[2.5cm][r]{NGC2253}    & 2.45 \\ 
NGC1093                         & 2.71 \\ 
\makebox[2.65cm][r]{NGC5947*}   & 2.58 \\
\makebox[2.5cm][r]{NGC6004}     & 2.46 \\ 
\makebox[2.5cm][r]{NGC2253}     & 2.45 \\ 
\makebox[2.5cm][r]{NGC2540}     & 2.68 \\ 
NGC2410                         & 2.80 \\ 
\makebox[2.5cm][r]{NGC5522}     & 2.73 \\ 
NGC2639                         & 2.70 \\
\makebox[2.5cm][r]{NGC0160}     & 2.84 \\ 
NGC2906                         & 2.39 \\
\makebox[2.65cm][r]{NGC0001*}   & 2.54 \\ 
\makebox[2.5cm][r]{NGC6063}     & 2.48 \\ 
\makebox[2.6cm][r]{UGC09777}    & 2.47 \\ 
NGC2916                         & 2.73 \\ 
\makebox[2.65cm][r]{NGC0001*}   & 2.54 \\ 
NGC6394                         & 2.90 \\ 
\makebox[2.60cm][r]{UGC12810}   & 2.93 \\
NGC7311                         & 2.71 \\ 
\makebox[2.5cm][r]{NGC0160}     & 2.84 \\ 
NGC7466                         & 2.85 \\ 
\makebox[2.65cm][r]{NGC2596*}   & 2.92 \\ 
\makebox[2.5cm][r]{NGC5980}     & 2.71 \\
UGC00005                        & 2.92 \\ 
\makebox[2.65cm][r]{NGC2596*}   & 2.92 \\ 
\makebox[2.5cm][r]{NGC5980}     & 2.71 \\ 
\makebox[2.65cm][r]{NGC0001*}   & 2.54 \\ 
UGC03973 $^{\dagger}$           & 2.38 \\
\makebox[2.75cm][r]{UGC02311*}  & 2.57 \\ 
\makebox[2.5cm][r]{NGC6032}     & 2.55 \\ 
\hline
\end{tabular}

\caption[j$_{*}$ values]{(1) Galaxy name as in Table \ref{tab:Sample}, (2) j$_{*}$ in the region dominated by the disc.  
}
\label{tab:Sample_j} 
\end{center}
\end{table}

\end{appendix}

\end{document}